\documentstyle[aaspp4]{article}
\begin{document}
\title{Dynamics of the  star S0-1 and the nature
of the compact dark object at the Galactic center}
\author{F. Munyaneza, D. Tsiklauri\footnote{permanent address:
Physics Department, Tbilisi State University, 3 Chavchavadze Ave.,
380028 Tbilisi, Georgia},
and R.D. Viollier}
\affil{Physics Department, University of Cape Town, Rondebosch 7701,
South Africa}
\begin{abstract}
It has been recently shown (Munyaneza, Tsiklauri and Viollier 1998)
that the analysis  of the orbits
of the fast moving stars close
to Sgr A$^{*}$ (Eckart and Genzel 1997) provides
a valuable tool to probe
the gravitational potential near the Galactic center.
As an example, we present here the results on a calculation
of possible orbits of the star S0-1 in both, the black hole
 and degenerate neutrino ball scenarios of the central mass, based
on the   recent measurements of
stellar proper motions at the Galactic center by
Ghez et al. 1998. Taking into account
the error bars of their analysis,
it is shown  that within  a few years time,
the orbit of S0-1 may indeed
reveal the nature of the supermassive compact dark object at the
Galactic center.
\end{abstract}

\section{Introduction}

The determination of the mass distribution near the center of our Galaxy
  and the question, whether
it harbours a supermassive black hole (BH) or not,
 have been long-standing issues
(Oort 1977; Genzel and Townes 1987;
  Genzel et al. 1994 and Ho 1998 for a  recent review).
Various  techniques have been used
to find the mass of this  supermassive
compact dark object which is  usually
identified with  the radio source Sagittarius A*
(Sgr A$^*$) at or near  the Galactic center.
The most detailed information to date comes from the
 statistical analysis  of the dynamics of  stars
moving  in the
 gravitational field
of  the central mass distribution (Sellgren et al. 1987;
Rieke and  Rieke 1988; McGinn et al. 1989; Sellgren et al. 1990;
Lindqvist et al. 1992;
Haller et al. 1996; Eckart and Genzel 1996; Genzel et al. 1996;
Eckart and Genzel 1997;  Genzel et al. 1997; Ghez
et al. 1998).
Genzel et al. 1997 have established that the central dark object
has a mass of $ (2.61 \pm0.76) \times 10^{6} M_{\odot}$,
concentrated  within a radius of 0.016 pc and located
very close to Sgr A$^*$.
In the most recent observations,
 Ghez at al. 1998 confirm a mass of
 $(2.6 \pm 0.2) \times 10^{6}M_{\odot}$,
enclosed within a radius of 0.015 pc. In the latter observations,
the accuracy
of the  velocity measurements  in the central  arcsec$^{2}$ has been
improved considerably,
and thus the error bar on the central mass has
been reduced by about a factor of 4.
In both  data sets, the presence of a supermassive compact dark object
is revealed by the fact that several stars
 are moving within a projected distance
of less than 0.01 pc from the central radio source Sgr A$^*$
at projected  velocities in excess of 1000 km/s.

For completeness, we mention here that
 the mass distribution at the Galactic
center
 could also be studied through the motion
of gas clouds and streamers (Lacy et al. 1980;
Genzel \& Townes 1987; Lacy et al. 1991).
However, gas flows may be easily perturbed by non-gravitational
forces such as shocks, radiation pressure, winds, magnetic fields,etc.,
 and hence this probe is considered
to be  less reliable
for  determining  the mass of the compact dark object
at the Galactic center.

The non-thermal spectrum of Sgr A$^*$ (Serabyn et al. 1997),
that has been shown to originate from a very compact source
 (Rogers et al. 1994; Genzel et al. 1997;
Ghez et al. 1998),
and the low proper motion of Sgr A$^{*}$ (Backer 1996)
have led many (e.g. Lynden-Bell and Rees 1971) to suggest
that Sgr A$^*$ may be a supermassive BH of mass
 $\sim 2.6 \times 10^{6}M_{\odot}$.
Supermassive  BHs have also been inferred
 for several other galaxies
such as M87 (Ford et al. 1994; Harms et al. 1994; Macchetto et al. 1997) and
NGC 4258 (Greenhill et al. 1995; Myoshi et al. 1995).
Taking this suggestion seriously, one is immediately faced  with
fundamental issues such as the
 prevalence of supermassive BHs in
the nuclei of normal galaxies and  the nature of
the accretion mechanism that makes Sgr A$^*$ so much
fainter than typical active galactic nuclei (Melia 1994; Narayan et al. 1995).
However, as the best current observations probe
the gravitational potential at  radii
$4\times 10^{4}$ larger than the Schwarzschild radius
of a BH of mass
$2.6\times10^{6}M_{\odot}$ (Ghez et al. 1998),
it is perhaps  prudent not to focus  too much on the BH scenario,
without having explored
alternative scenarios for the supermassive compact
dark object.

One alternative to the BH scenario is a very
compact stellar cluster
(Haller et al. 1996, Sanders, 1992).
However, based on the evaporation and collision time
stability criteria, it is doubtful that such clusters could have
survived up to the present time
(see Moffat 1997 for an alternative point of view). Indeed,
in the case of our Galaxy and NGC 4258,
Maoz (1995,1998) has found that even
the lower limits to the half-mass densities
of such compact clusters  ($1 \times 10^{12}M_{\odot}{\rm pc}^{-3}$ for
NGC 4258 and $6 \times 10^{11}M_{\odot}{\rm pc}^{-3}$ for our Galaxy)
are too large that they could
be due to stable clusters of stellar
or substellar remnants.
The estimated maximal lifetimes for such  dense clusters
are about $10^{8}$ years for our  Galaxy and a few $10^8$ years for the
NGC 4258, i.e. much shorter than the age of the Universe.
This seems to rule out the existence of dense clusters at the
centers of the above mentioned galaxies,
unless we are prepared to believe that we happen to live
in a privileged epoch of the lifetime of the Universe.
Note, however, that for other galaxies, such as M31, M32, M87, NGC 3115,
NGC 3377, NGC 4261, NGC 4342, NGC 4486B and NGC 4594,
maximal lifetimes of dense stellar clusters are
in excess of $10^{11}$ years.
Moreover, it should be acknowledged that
the uncertainties in the understanding
of the core collapse process of such dense clusters
still leave some room for speculation about a
possible interpretation of the supermassive compact
dark objects at the centers of galaxies
(including both, our Galaxy and NGC 4258) in terms
of e.g. core-collapsed clusters (Maoz 1998).
But, apart from a cluster of very low mass BH's
that is free of stability problems, the most
attractive alternative to a dense stellar cluster
is a cluster of elementary particles.

In fact, in the  recent past, an alternative model for the
 supermassive compact dark  objects in galactic centers
has been  developed
(Viollier et al. 1992, 1993; Viollier 1994;
Tsiklauri  and Viollier 1996, 1998a,b, 1999;
Bili\'c et al. 1998; Bili\'c et al. 1999).
The cornerstone  of this model
is that the dark matter at the center
of galaxies is made of nonbaryonic matter in the form
of massive neutrinos that interact  gravitationally forming
 supermassive neutrino balls in which the degeneracy
pressure of the neutrinos  balances their self-gravity.
Such neutrino balls could have been formed in the early
Universe during  a first-order gravitational
phase transition (Bili\'c and Viollier 1997,1998,1999a,b).
In fact, it has been recently shown that the dark matter concentration
observed through stellar motion
at the Galactic center (Eckart \& Genzel 1997; Genzel et al. 1996)
is consistent with a supermassive object of $2.5 \times 10^{6}$
solar masses  made of self-gravitating, degenerate heavy neutrino
matter (Tsiklauri \& Viollier 1998a).
Moreover, it has been shown  that  an
acceptable fit to the infrared and
 radio
spectrum above 20 GHz , which is presumably
 emitted by the compact dark object,
  can be reproduced
in the framework of  standard
accretion  disk theory (Tsiklauri \& Viollier 1999; Bili\'c et al. 1998), in
terms of a baryonic disk  immersed in the shallow potential of
 the degenerate neutrino ball of $2.5 \times 10^{6}$ solar masses.

The purpose of this paper is to compare
the predictions
of these two models for the supermassive
 compact dark object at the center
our Galaxy, i.e.
(i) the black hole scenario and
(ii) the degenerate neutrino ball scenario as an example
 of an extended object.
Both these models are not in contradiction
with the technologically
challenging proper motions observations and
their statistical interpretation
 (Genzel et al. 1997; Ghez et al. 1998) .
It is therefore desirable to have
an additional independent dynamical test,
in order to distinguish between these two possible  scenarios
 describing the compact dark object at the center of our
Galaxy.
In the recent past, mainly statistical arguments involving
many stars have been used  to determine
the gravitational potential
at the Galactic center.
However, in this paper, we would like to
demonstrate that it is
also possible
to draw definite conclusions
from the motion of individual stars,
in particular, in the immediate vicinity of the Galactic center,
where statistical arguments cannot be easily applied
due to the low density of stars.
To this end, we have  recently calculated the orbits
(Munyaneza , Tsiklauri and Viollier 1998) of the fastest moving infrared
source S1 using the Genzel et al. 1997 data for a  supermassive BH
or a neutrino ball mass of $2.61 \times 10^{6}$ solar masses.
We have shown that tracking  the orbits of S1
offers a good opportunity to distinguish
in a few years time between  the two scenarios
for the supermassive compact dark object.
Here we perform
a full analysis of the
orbits of the same star S0-1 based
on  the most recent Ghez et al. 1998 data,
including all  the error bars of the measurements.
A  distance to the Galactic center  of 8 kpc has been assumed
throughout this paper.

This paper is organized as follows:
In section~2, we present the equations
that describe degenerate neutrino balls and
we establish some constraints on the neutrino mass based on
Ghez et al. 1998 data.
In section~3, we study the dynamics of S0-1
and conclude with the discussion in section~4.

\section{The compact dark  object as a neutrino ball}

Dark matter at the Galactic center
can be described by the gravitational potential $\Phi(r)$
of the neutrinos and antineutrinos that satisfies
Poisson's equation
\begin{equation}
\label{eq:01}
\Delta \Phi =4\pi G \rho_{\nu},
\end{equation}
where $G$ is  Newton's gravitational
constant and  $\rho_{\nu}$ is the 
mass density of the neutrinos and antineutrinos.
Neutrino matter will interact gravitationally to form
supermassive neutrino balls in which
self-gravity of the  neutrinos is being balanced
by their degeneracy pressure $P_{\nu}(r)$ according to the
 equation of hydrostatic equilibrium
\begin{equation}
\label{eq:02}
\frac{dP_{\nu}}{dr}=-\rho_{\nu}\frac{d\Phi}{dr} \ .
\end{equation}
In order to solve equation (\ref{eq:01}) , one needs a relation
between the pressure $P_{\nu}$ and the density $\rho_{\nu}$.
To this end we choose
 the polytropic equation of state of  degenerate neutrino matter, i.e.
\begin{equation}
\label{eq:03}
P_{\nu}=K\rho_{\nu}^{5/3},
\end{equation}
where the polytropic constant $K$  is given by (Viollier, 1994)
\begin{equation}
\label{eq:04}
K=\left(\frac{6}{g_{\nu}}\right)^{2/3}\frac{\pi^{4/3}\hbar^{2}}
{5m_{\nu}^{8/3}}.
\end{equation}
Here, $m_{\nu}$ denotes the neutrino mass, $g_{\nu}$ is
 the spin degeneracy
factor of the neutrinos and antineutrinos, i.e. $g_{\nu}=2$ for
 Majorana and $g_{\nu}=4$ for Dirac neutrinos and antineutrinos.
We now introduce  the dimensionless potential  and
radial variable, $v$ and  $x$ , by
\begin{equation}
\label{eq:05}
\Phi(r) = \frac{GM_{\odot}}{a_{\nu}}\left(v'(x_{0})-\frac{v(x)}{x}\right),
\end{equation}
\begin{equation}
\label{eq:06}
r=a_{\nu}x,
\end{equation}
where $x_{0}$ is the  dimensionless radius of the neutrino ball,
and the scale factor $a_{\nu}$ which plays here the role of a 
length unit is given by
\begin{equation}
\label{eq:07}
a_{\nu}=2.1376~{\rm lyr} \times
\left(\frac{17.2 \  {\rm keV}}{m_{\nu}c^{2}}\right)^{8/3}g_{\nu}^{-2/3}.
\end{equation}
Assuming
spherical symmetry, we finally arrive at the non-linear
 Lan\'{e}-Emden equation
\begin{equation}
\label{eq:08}
\frac{d^{2}v}{dx^{2}}=-\frac{v^{3/2}}{x^{1/2}},
\end{equation}
with polytropic index 3/2. The boundary conditions are chosen
in such a way that $v$ vanishes
 at the boundary $x_0$ of the neutrino ball. The mass $M_{B}$ of
a (pointlike) baryonic star at the center of the neutrino ball
is fixed by
 $v(0)= M_{B}/M_{\odot}$.
  The case  $M_B=0$ corresponds
to a pure neutrino ball without a pointlike source at the center.
The mass enclosed within a radius $r$
in a pure neutrino ball can be written in terms of
$v(x)$ and its derivative $v'(x)$ as
\begin{equation}
\label{eq:09}
M(r)=\int_{0}^{r}4\pi \rho_{\nu}r^{2}dr
 =-M_{\odot} \left(v'(x)x-v(x) \right).
\end{equation}

In order to describe  the compact dark object at the
Galactic center as a neutrino ball
and  constrain its physical parameters appropriately,
 it is worthwhile
to use the most recent observational data by Ghez et al. 1998, who
established that the  mass enclosed within 0.015 pc at the Galactic center
 is $(2.6 \pm 0.2) \times 10^{6}$ solar masses.
Following the analysis of Tsiklauri \& Viollier 1998a,
we choose the minimal neutrino   mass $m_{\nu}$
  to reproduce
the observed matter distribution, as
 can be seen from Fig.~1, 
where we have added the Ghez et al. 1998  and Genzel
et al. 1997  data points
with error bars.
In  Fig.~1 we include only the neutrino ball
contribution to the enclosed mass,
 as the  stellar cluster
contribution  is negligible by orders of
magnitude at these radii.
For a $M=2.4 \times 10^{6}M_{\odot}$ neutrino ball,
the constraints on the neutrino mass are
$m_{\nu} \ge 17.50$~ keV$/c^{2}$
for $g_{\nu} =2$  and $m_{\nu} \ge 14.72~$keV$/c^2$
for $g_{\nu}= 4$, and the radius of the neutrino
 ball is $R \le 1.50\times 10^{-2}~{\rm pc}$.
Using the value of $M=2.6 \times 10^{6}M_{\odot}$, the bounds on
the neutrino mass  are $m_{\nu} \ge 15.92$ keV$/c^{2}$ for
$g_{\nu}=2$ or $m_{\nu} \ge 13.39$ keV$/c^{2}$  for $g_{\nu}=4$
and the radius of the neutrino ball turns out to be
$R \le 1.88 \times 10^{-2} {\rm pc}$ .
Finally, for a $M=2.8 \times 10^{6}M_{\odot}$ neutrino ball,
the range of neutrino mass is
$m_{\nu} \ge 15.31 \  $keV$/c^{2}$  for $g_{\nu}= 2$ and
$m_{\nu} \ge  12.87 \  $keV$/c^{2}$ for $g_{\nu}=4$
and the corresponding neutrino ball 
radius $R \le 2.04 \times 10^{-2}$ pc.

\section{Dynamics of S0-1}

We investigate the motion of S0-1 that is the star  closest to
the Galactic center, and at the same time, also
 the  fastest  of the 15 stars in the central
 arcsec$^2$ around Sgr A$^*$.
We study the motion of S0-1
in the gravitational potential near  Sgr A$^*$,
assuming that the central object
is either a  BH of mass $M$ or
a spatially extended object represented by
a neutrino ball of  mass
$M$, that  consists of self-gravitating
degenerate heavy neutrino matter.
 The BH or neutrino ball mass $M$ will
be taken to be 2.4, 2.6 and 2.8 $\times 10^{6}$ solar
 masses  which corresponds to the range allowed
 by the Ghez et al. 1998 data.
We use  Newtonian dynamics,
as the problem  is essentially nonrelativistic,
because the mass of the neutrino ball
is much less than the Oppenheimer-Volkoff limit corresponding
to this particular  neutrino mass (Bili\'c, Munyaneza \& Viollier 1999).
Consequently, we can write  Newton's equations of motion
as
\begin{equation}
\ddot{x}=-\frac{GM(r)}{(x^{2}+y^{2}+z^{2})^{3/2}}x,
\label{eq:10}
\end{equation}
\begin{equation}
\ddot{y}=-\frac{GM(r)}{(x^{2}+y^{2}+z^{2})^{3/2}}y,
\label{eq:11}
\end{equation}
\begin{equation}
\ddot{z}=-\frac{GM(r)}{(x^{2}+y^{2}+z^{2})^{3/2}}z,
\label{eq:12}
\end{equation}
where $x$, $y$, $z$ denote the  components of the radius vector of the
star S0-1 and $r=\sqrt{x^2+y^2+z^2}$, Sgr A$^*$ being the origin
of the coordinate system.
We thus assume that the center of the neutrino ball
and the  BH is at the position of Sgr A$^*$.
The dot denotes  of course the derivative with
respect to time. In the case of a BH,
 $M(r)=M$ is independent of $r$,
while in the  neutrino ball scenario,
$M(r)$ is given by Eq.~(\ref{eq:09})
and it reaches $M(R)=M$ at the radius of the neutrino ball $R$.
The initial  positions and velocities for this system
of equations  are taken
to be those of S0-1 in 1995.4, when
 the coordinates of S0-1 were
${\rm RA}=-0.107''$ and ${\rm DEC}=0.039''$.
The $x$ and $y$ components of the projected velocity are
$v_{x}=470 \pm 130$~km/s
 and $v_{y}=-1330 \pm 140$~km/s
(Ghez et al. 1998), respectively.
Here $x$ is  opposite to the RA direction and $y$ is in the DEC direction.

In Fig.~2 we plot two typical orbits of S0-1 corresponding to a
BH and neutrino ball  mass of  $M=2.6 \times 10^{6} M_{\odot}$.
The input  values for $v_{x}$ and $v_{y}$ are 470 km/s and -1330 km/s,
respectively. The {\it z}-coordinate of the star S0-1
is assumed to be zero and the velocity component
in the line-of-sight of the star S0-1, $v_{z}$,  has also
been set equal to zero in this graph. The filled square
labels denote the time in years from 1990 till 2015.
In the case of a BH, the orbit of S0-1 is an ellipse,
with Sgr A$^*$ being located in one focus
(denoted by the star in the figure). The period of S0-1
is 12.7 years  and the minimal and maximal distances
from Sgr A$^*$ are 1.49 and 7.18 light days, respectively.
In the case of a neutrino ball, the
orbit will be bound but not closed,
with minimal and maximal
distances from Sgr A$^*$ of 3.98 and 42.07 light days, respectively.
It can be seen from Fig.~2 that,
 in the case of a neutrino ball,
S0-1 is deflected much less than for a BH,
as the gravitational force at a given distance from Sgr A$^*$ is determined
by the mass enclosed within this distance. Using Eq. (\ref{eq:09})
we can estimate
 the mass enclosed within a radius
corresponding to the projected  distance of
S0-1 from Sgr A$^*$ ($4.41 \times 10^{-3}$ pc ) to be
$\sim 1.8 \times 10^{5} M_{\odot}$.
Thus, in the case of a neutrino ball, the force acting on S0-1
 is  about 14 times less than in the case of a BH.
This graph can serve to establish, whether  Sgr A$^*$ is a BH
or an extended object, due to the fact that the
 positions of S0-1 will differ as time goes on in the two scenarios.
However, this conclusion is perhaps too optimistic,
 as we have not yet  considered
 (i) the uncertainties in $v_{x}$ and $v_{y}$,
 (ii) the error bars  in the total mass of the BH or
neutrino ball,
(iii) the complete lack of information on $z$ and $v_{z}$.

As a next step, we investigate the dependence of the orbits on
the uncertainties in the velocity components.
The results of this calculation are presented
in Fig.~3 where we  have set $z=v_{z}=0$.
In the case of a BH, the
  orbits of S0-1 are ellipses ,
 while the other 5 thick lines are bound orbits of S0-1
for the neutrino ball scenario. The spread of the orbits
induced by the error bars in $v_{x}$ and $v_{y}$ is small
compared to that of the recent analysis 
based on the  Genzel et al. 1997 data
 (Munyaneza, Tsiklauri \& Viollier 1998).
 The time labels, represented by filled squares on the orbits,
are placed in intervals of 5 years: starting  from 1995.4
up  to 2005 in the
case of a BH,
and up to 2015 in the case of a neutrino ball.
The periods of S0-1 for different  orbits
 vary between
10 and 17 years for the BH scenario.
We  thus see that the error bars in $v_{x}$ and $v_{y}$ do not alter
the  predictions  of Fig.~1 in substance. We now would like to study,
 how the orbits are changed if we let 
 the mass of the neutrino
ball or the BH vary within the estimated error bars (Ghez et al. 1998).

In Fig.~4 and 5, we present the results of our calculations,
for both scenarios, with central  masses of $M=2.4 \times 10^{6} M_{\odot}$
and $M=2.8 \times 10^{6} M_{\odot}$, respectively.
The neutrino masses consistent
with the Ghez et al. (1998) data are are $m_{\nu} \ge 17.50$ keV$/c^{2}$ for
a  $M=2.4 \times 10^{6} M_{\odot}$  neutrino ball  and
$m_{\nu} \ge 15.31$  keV$/c^{2}$ for
 a $M=2.8 \times 10^{6} M_{\odot}$ neutrino ball.
The filled squares represent the time labels
spaced by 5 year intervals as in Fig.~3.
In the BH scenario, the
 periods of S0-1
with  $M= 2.4 \times 10^{6}M_{\odot}$   vary between 11
years and 20 years,
while in the case of a $M=2.8 \times 10^{6}M_{\odot}$,
the periods vary
between 9.5 and 15 years.
Comparing the orbits of S0-1 in Fig.~4 and 5 with those
in Fig.~3, we conclude that  the errors bars in the
  total mass of the BH
 or neutrino ball
make no qualitative difference for the motion of S0-1.
In both scenarios of the supermassive compact dark object,
all the orbits considered
for three different values of the BH or neutrino ball
mass are bound for $z=v_{z}=0$,
as can be seen from Fig.~6 and 7, where
  the escape and circular velocities
are plotted as functions of the distance from Sgr A$^*$.  In these graphs,
we have also included the Ghez  et al. (1998) data
with error bars, for the 15 stars in the central  arsec$^2$,
assuming that the velocity component and distance from
Sgr A$^*$  in the line-of-sight are both zero, i.e. $v_z = 0$ and
$z=0$.
Thus, the data points
are lower bounds on the true circular or escape velocity and radius, and
the real values lie in the quarter-plane to the right-and-up
of the measured data point.
For instance, the innermost data point describing the star
S0-1 is in both scenarios, consistent with a bound orbit if $|z|$
and $|v_{z}|$ are not too large, as can be seen from the escape velocity
in Fig.~6.
However, S0-1 cannot be interpreted as a virialized star in the neutrino
ball scenario, as is evident from the plot of the circular velocity
in Fig~7; it thus would have to be an intruder star.
If the projected velocity of a star
at a given projected distance from Sgr A$^*$ is larger than the
escape velocity
at the same distance (assuming $z=0$), the neutrino ball
scenario is virtually ruled out, since the kinetic energy
of the star would have to be very large at infinity.

We now turn to the investigation
of the dependence of the orbits on
 the $z$ coordinate  and $z$ component of the velocity of the star
S0-1. The two quantities,  $z$ and $v_{z}$, are the major source
of uncertainty in determining the exact orbit of the star S0-1.
However, this shortcoming will not substantially
affect
the predictive power of our model,
as we will see below.
In Fig.~8 we show the results of a calculation
of the  dependence of the orbit
 on $z$ for a $M=2.6\times 10^{6}M_{\odot}$ neutrino
ball or BH. The input values for $v_{x}$ and $v_{y}$
are  fixed at 470 km/s and  -1330  km/s, respectively, and
$v_{z}$ is assumed to be zero. The $z$-coordinate is varied from zero
up to the   radius of the neutrino ball, i.e. the distance from
Sgr A$^*$ beyond which there is obviously no difference between
the BH and the neutrino ball scenarios.
In this case, the radius of the neutrino ball
$ 1.88 \times 10^{-2}$ pc  or 0.485 $''$.
The top panel represents the orbits in the case of
a BH, for different values of $z$, while the lower
panel describes the  dependence of the orbit  on $z$ in the neutrino ball
scenario.
We conclude from this plot that, increasing  $|z|$ has
the effect of shifting the orbits towards the lower right corner
of the graph. This is, obviously, due to the fact that increasing
 $|z|$ means going further away from the scattering center, thus
yielding less deflection of the orbit.
Moreover, in the neutrino ball scenario,
 the dependence on $z$ is  relatively
insignificant, as long as $|z|$ is smaller
than the radius of the neutrino ball.
This is in accordance with the fact that
for small distances from the center, the potential of
a neutrino ball   can be approximated by
a harmonic  oscillator-type potential,
where the Newtonian equations of motion decouple
in Cartesian coordinates.
The dependence of the orbits of S0-1 on $v_{z}$ has a similar effect
as in the previous graph.Here,  we have fixed $z$ to zero
and $v_{z}$ has been varied as an input parameter.
Increasing $|v_{z}|$ yields  a greater velocity of the star and,
 obviously,  a fast moving star will be deflected
less than a star with smaller $|v_{z}|$.
The results of this calculation are summarized
in Fig.~9.

\section{Conclusion and discussion}

We have demonstrated that the orbits of S0-1
differ substantially for  the BH and neutrino ball
scenarios of  the Galactic center, especially
with the new Ghez et al. (1998) data.
We have shown that using  these  data,
the error bars in velocities of S0-1 and mass of the central object
do not change the pattern of the orbits of S0-1. 
In the case of a BH, the orbit
 of S0-1 is much more curved  than in the neutrino ball scenario, as
long as $|z|$ is smaller than the radius of the neutrino ball.
Increasing  $|z|$ and $|v_{z}|$ shifts the orbits to the  lower right
corner of the graph, and this gives us a key to
establish the allowed regions of  S0-1 depending on
whether it is  a BH or a neutrino ball,
{\it irrespective} of the values of the parameters $z$ and $v_{z}$.
In Fig.~ 10   three orbits are plotted:
the upper-leftmost orbit of S0-1 corresponding
to the neutrino ball scenario (actually, line 9 of Fig. 4) and
two orbits in a BH scenario with the smallest
minimal and maximal distances from Sgr A$^*$ (ellipses 2 and 4 from Fig.5).
This  figure serves as a test to distinguish the supermassive BH
scenario from the neutrino  ball model of the Galactic
center. It is clear that,  as the observations proceed within
the next   year, one might  be able to tell the difference between
the two models of the supermassive compact dark object at
the center of our Galaxy.

 If the star is found
in the region $F$  inside
the ellipses, this will rule out both the BH and the neutrino ball
scenario of Sgr A$^*$, as seen in Fig.~10. We can estimate the minimal
distance of approach to
 Sgr A$^*$  to be  0.909 light days.
If the orbit of S0-1 ends up  in the upper-left zone of the
thick line, this will clearly rule out the neutrino ball  scenario
 for the chosen
lower limit of the neutrino mass. However, if S0-1 is found in the lower right
corner of the same line (i.e. below the thick line),
then the supermassive object can be interpreted as either a
neutrino ball or a BH with  a large $z$ or $v_{z}$ parameter.
One can of course repeat this analysis for several stars in the
central arcsec$^2$ and use some statistical
arguments: should there be no stars in the black hole
zone, and many stars found in the zone
for black holes
and neutrino balls, the black hole
interpretation of the supermassive compact dark object
at the Galactic center would become less
 attractive, as some of the stars should
be moving close to the plane perpendicular to the line-of-sight, i.e.
they should have small $|v_{z}|$ and $|z|$.

The neutrino masses used for the
 neutrino ball are lower limits.
We note that increasing  the neutrino mass
will make the radius smaller (the neutrino ball  radius
scales as $ \propto m_{\nu}^{-8/3}$) and, when it reaches the mass
corresponding to the
Oppenheimer-Volkoff limit, there
will be little difference between the two scenarios.

\section{Acknowledgements}

One of us (F. Munyaneza)  gratefully acknowledges
funding from the Deutscher Akademischer Austauschdienst and the
University of Cape Town. This work is supported by
the Foundation for Fundamental Research (FFR).

\newpage

\centerline{Figure captions:}

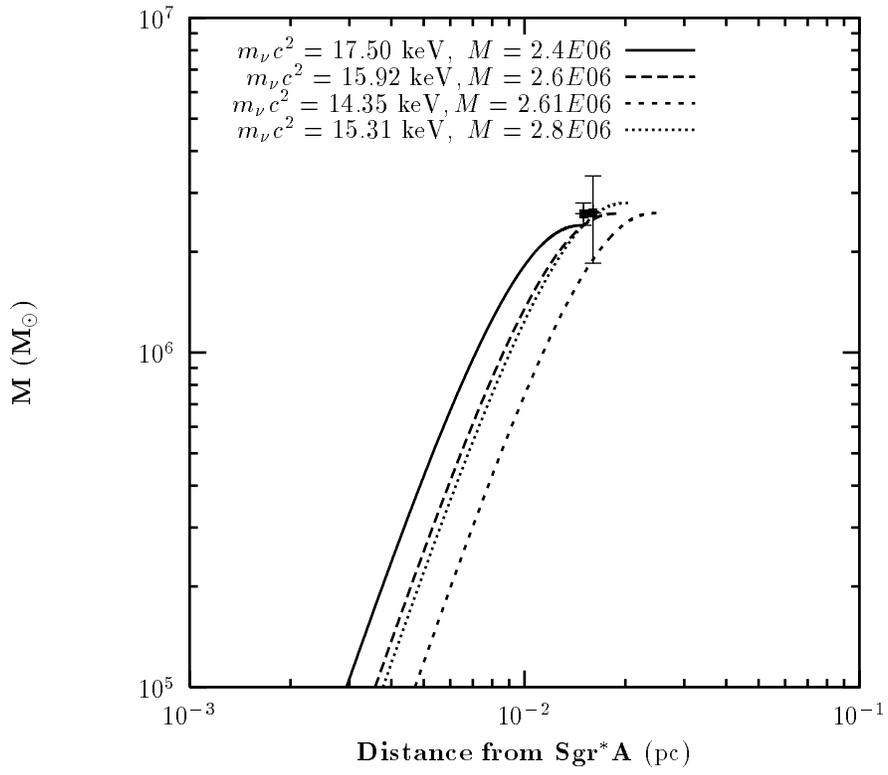
Fig.~1: The mass enclosed within a distance
of the center of  a  neutrino ball of
2.4, 2.6, 2.61 and 2.8  millions solar masses.
Based on the Ghez et al. 1998 data, the bounds on the neutrino mass
are $m_\nu \ge 17.50$ keV$/c^2$  for $g_{\nu}=2$  or
$m_{\nu} \ge  14.72$ keV$/c^2$ for $g_{\nu}=4$ and $M=2.4 \times
10^{6}M_{\odot}$.
For  $M=2.6  \times 10^{6} M_{\odot}$ the bounds on the neutrino mass
are $m_\nu \ge 15.92$ keV$/c^2$ for $g_{\nu}=2$ and
$m_{\nu}  \ge  13.39$ keV$/c^2$  for $g_{\nu}=4$.
Finally, a neutrino mass range of $m_{\nu} \ge  15.31$ keV$/c^{2}$
for $g_{\nu}=2$ or $m_\nu \ge 12.87$ keV$/c^2$
for $g_{\nu} =4$ is  consistent
with
a supermassive object of $M=2.8 \times 10^{6}M_{\odot}$.
The Ghez  et al. 1998 and Genzel
et al. 1997 data points with error bars are
also shown in this graph.

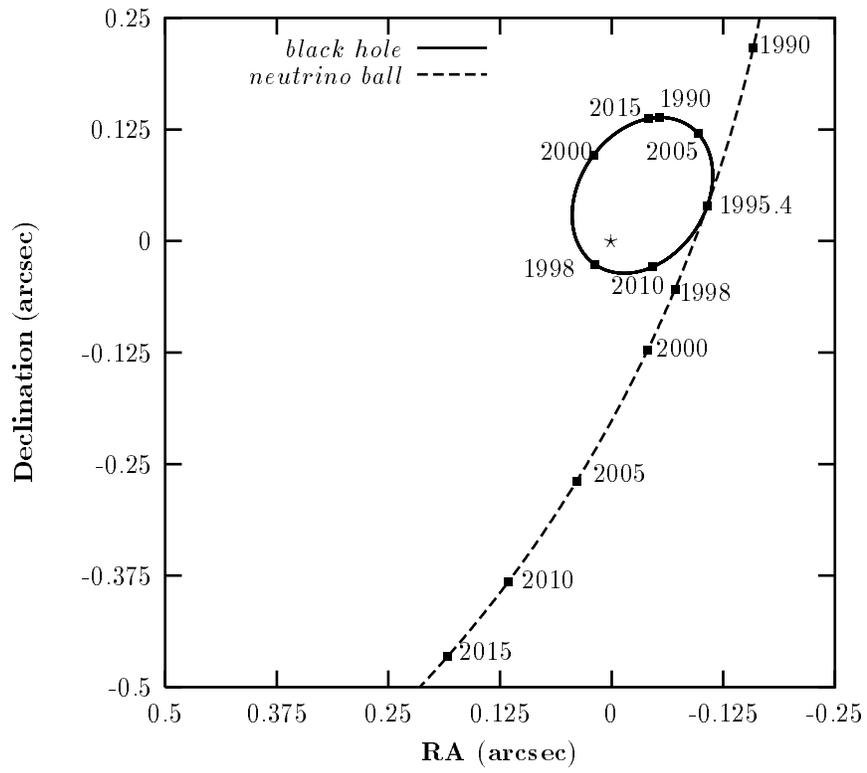
Fig.2: Projected orbits of the star S0-1 for
BH and neutrino ball
 scenarios with $M=2.6 \times 10^{6} M_{\odot}$
and  $v_z=z=0$. The velocity components of S0-1
are taken to be $v_{x}=470$ km/s and
$v_{y}=-1330$ km/s. The filled squares denote the time labels.
The period of S0-1 in a the BH scenario is 12.7 years
and the minimal and maximal distances from Sgr A$^*$ are 1.49 and
7.18 light days. The orbit of S0-1 in the neutrino ball
scenario is bound with minimal and maximal distances from Sgr A$^*$
of 3.98 and 42.07  light-days, respectively.

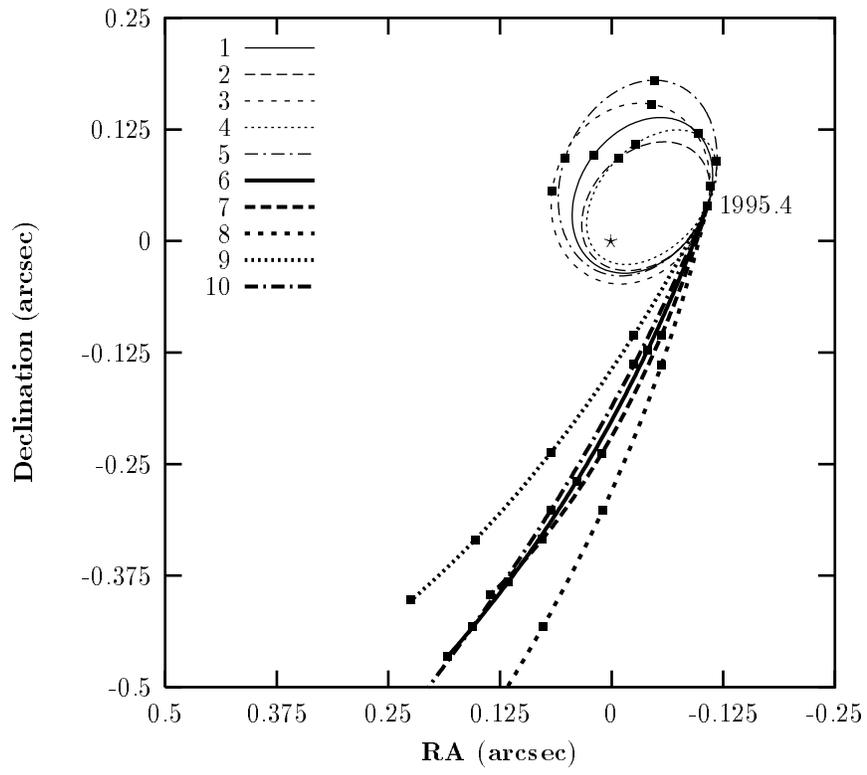
Fig.3: Projected orbits
of the star S0-1 in the case of a BH
or a neutrino ball of $M=2.6 \times 10^{6} M_{\odot}$ taking into
account the error bars in the velocity components.
The labels for the different orbits are:
1: $v_x=470$ km/s  and $v_y=-1330$ km/s (median
values),
2: $v_x=340$ km/s and $v_y=-1190$ km/s,
3: $v_x=340$ km/s and $v_y=-1470$ km/s,
4: $v_x=600$ km/s and $v_y=-1190$ km/s,
5: $v_x=600$ km/s and $v_y=-1470$ km/s.
 The periods of S0-1 for different orbits in the BH scenario
vary between  10 and 17 years.
The thick lines 6 to 10 correspond to the orbits
in the neutrino ball scenario with the following description:
 6: $v_x=470$ km/s  and $v_y=-1330$ km/s (median
values),
7: $v_x=340$ km/s and $v_y=-1190$ km/s,
8: $v_x=340$ km/s and $v_y=-1470$ km/s,
9: $v_x=600$ km/s and $v_y=-1190$ km/s,
10: $v_x=600$ km/s and $v_y=-1470$ km/s.
All the orbits in both scenarios  are bound.
The time labels (filled squares) on the orbits are placed
in intervals of 5 years, up to the year 2005 in the case of a
BH and up to 2015 in the case of a neutrino ball.

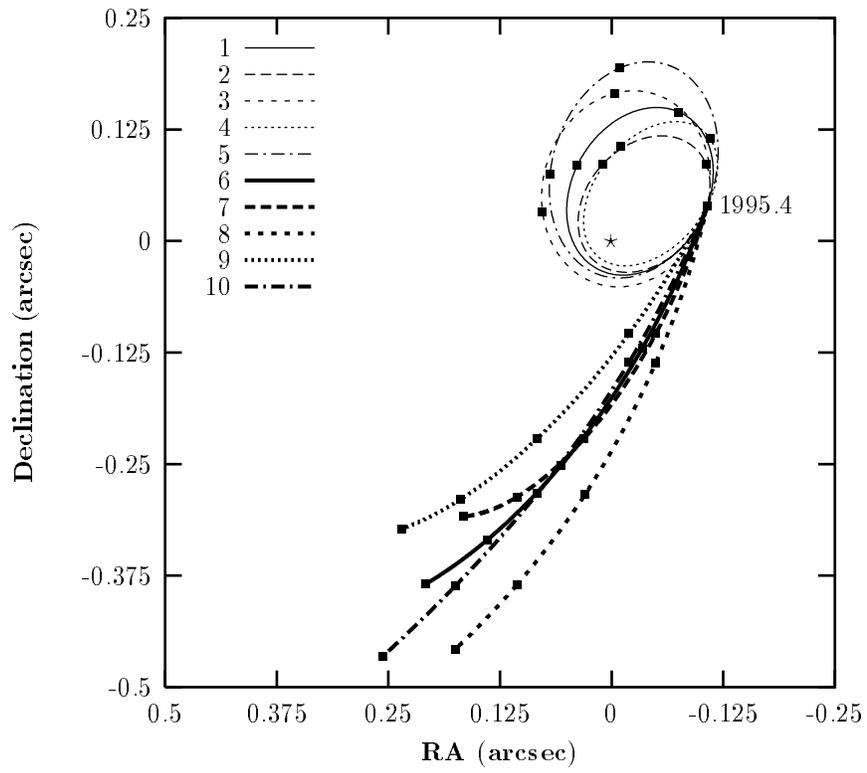
Fig.4: Projected orbits of the star S0-1 in the case
of a BH or neutrino ball with  $M=2.4 \times 10^{6}M_{\odot}$.
In this graph we explore how the orbits
are affected by the uncertainty in the mass of the
BH or neutrino ball. The orbits are calculated
for $z=v_{z}=0$ .
The description of the orbits are the same as  Fig.~3.
The periods of S0-1 in the BH scenario vary
between  11 and 20 years and all the orbits
are bound in both scenarios.

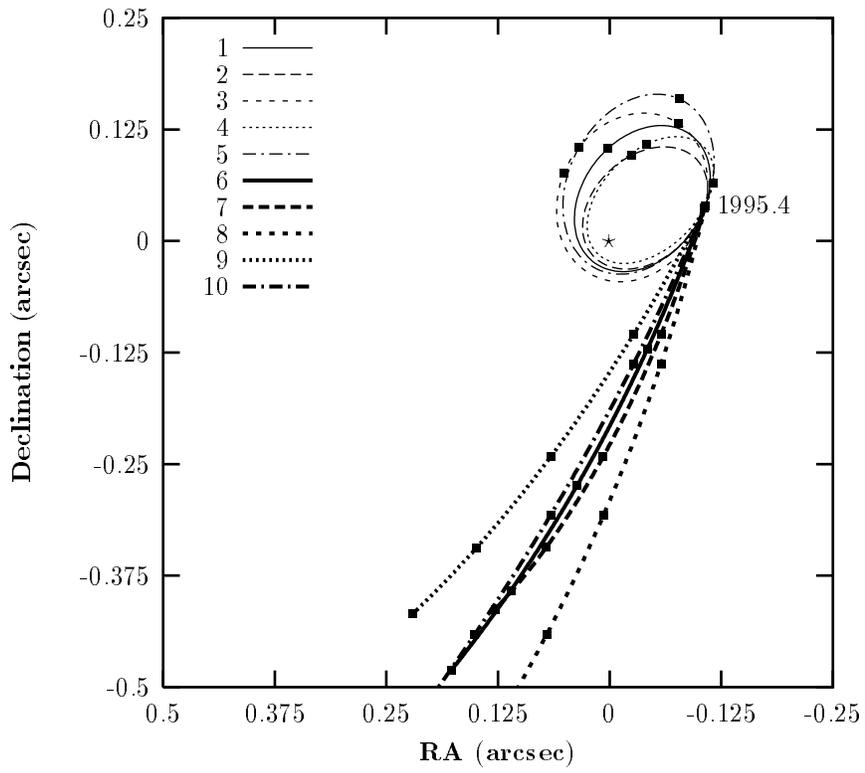
Fig.5: Projected orbits of the star S0-1  for  $M=2.8 \times 10^{6} M_{\odot}$.
All the orbits are bound and calculated for
different values  of the velocity components as in Fig.~3.
The periods of S0-1 vary between 9.5 years and 14.7 years in the BH
 scenario.

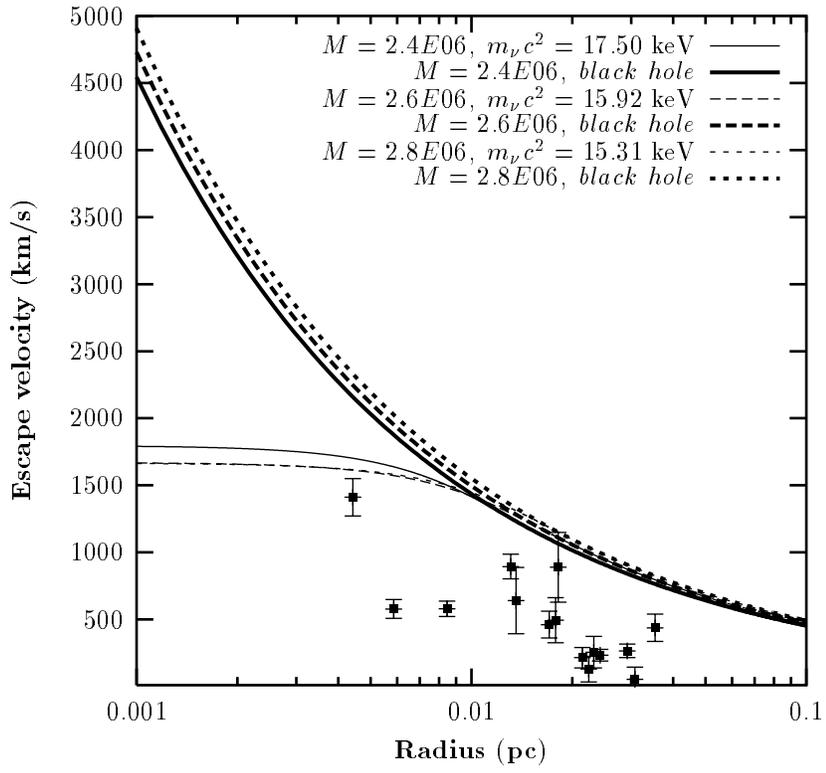
Fig.~6: The escape velocity as a function of the distance
from Sgr A$^*$ for BH and neutrino ball scenarios.
The value of the mass of the central object
 is varied as indicated on the graph.
The data points with error bars of
15 stars in the central  arcsec$^2$ are taken from Ghez et al. 1998
assuming that the projected velocity and distance
 from Sgr A$^*$ are equal to the  true velocity
and distance, respectively, i.e. $z=0$ and $v_z=0$. This graph shows that S0-1
is bound in both scenarios for different values of the mass
of the central object.

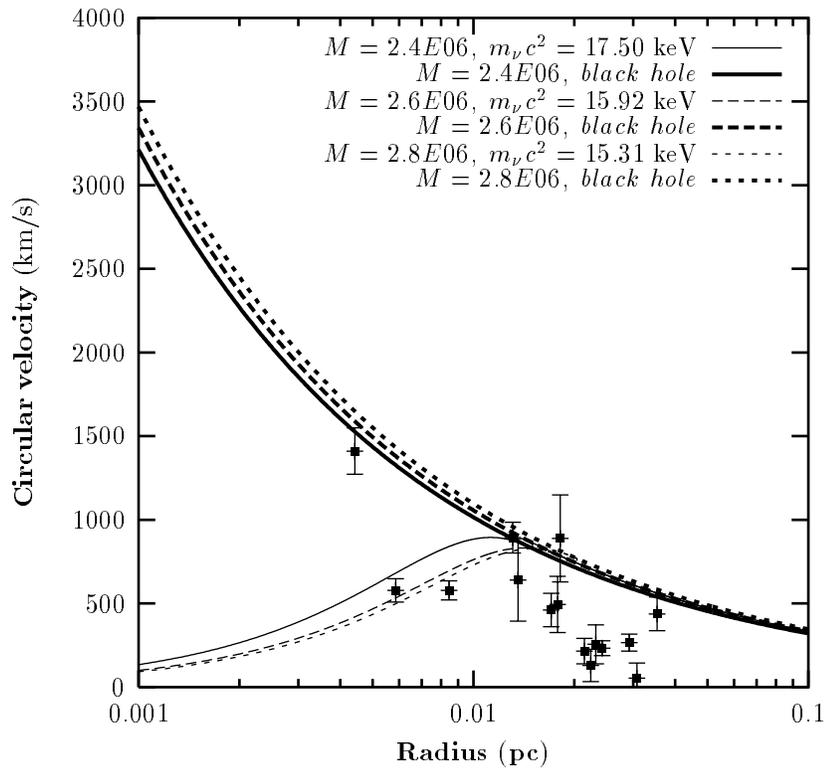
Fig.~7: The circular velocity as a function of the distance
from Sgr A$^*$ for BH and neutrino ball scenarios.
The  mass of the central object is
varied as indicated on the graph.
The data points with error bars of
15 stars in the central  arcsec$^2$ are taken from Ghez et al. 1998
assuming that the projected velocity and distance
 from Sgr A$^*$ are equal to the  true velocity
and distance, respectively, i.e. $z=0$ and $v_z=0$. This graph shows that
the orbits of S0-1
are almost circular in the case of the BH scenario
(see text for the discussion concerning the innermost data point).

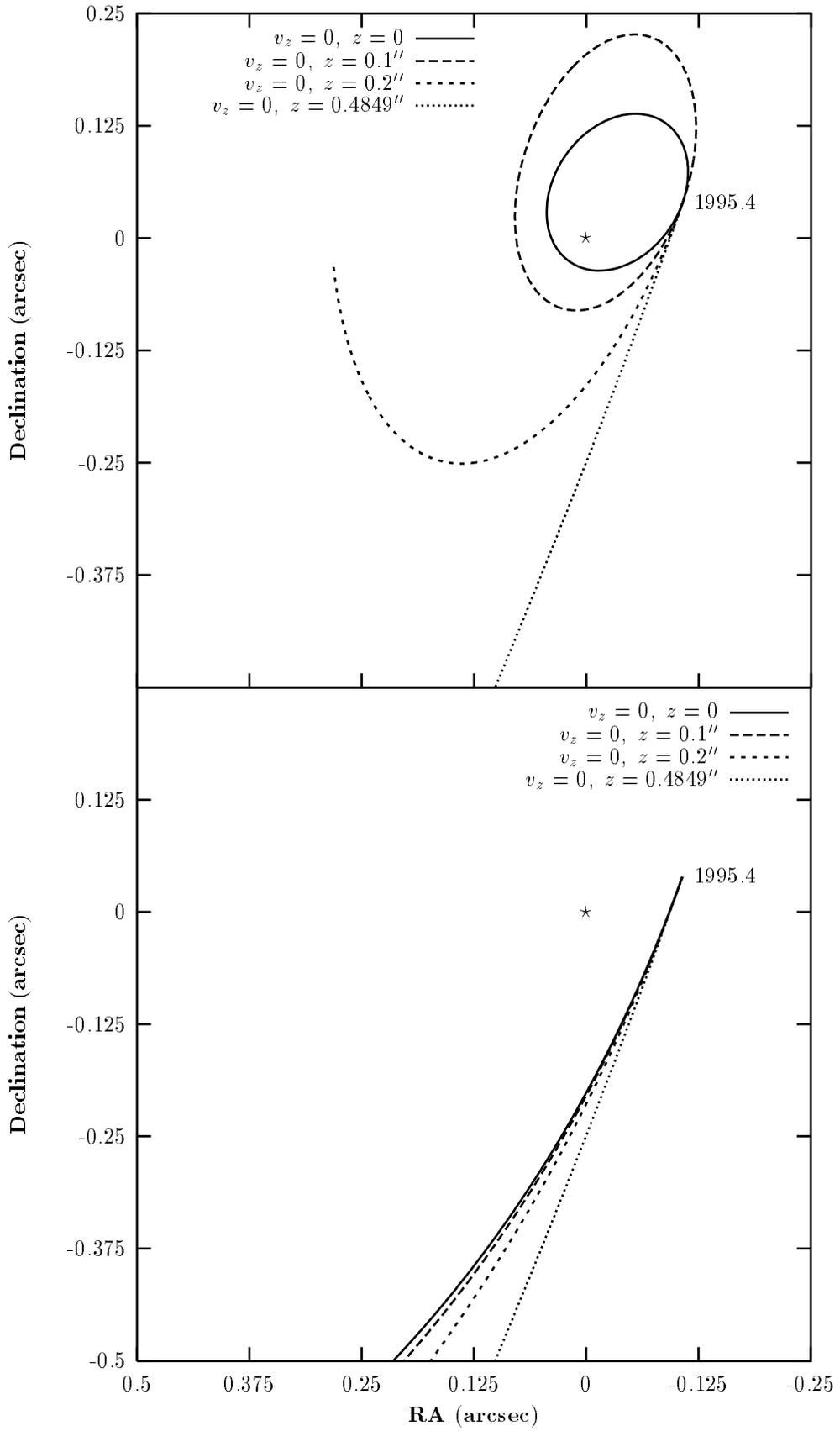
Fig.~8:
Projected orbits of the star S0-1 in the case
of a supermassive
BH (top panel)
and in the case of a neutrino ball (lower panel) with
 $M=2.6 \times 10^{6} M_{\odot}$.
In this graph we explore how the orbits  are affected by the
uncertainty in the $z$-parameter. The labels for the orbits are
given in the graph. Note, that for $z=0.4849''$, which
corresponds to the radius of the neutrino ball for the assumed
distance to the Galactic center, the orbits for a BH and
neutrino ball are identical, as it should be.
In this graph $v_x=470$ km/s, $v_y=-1330$ km/s and $v_z=0$.

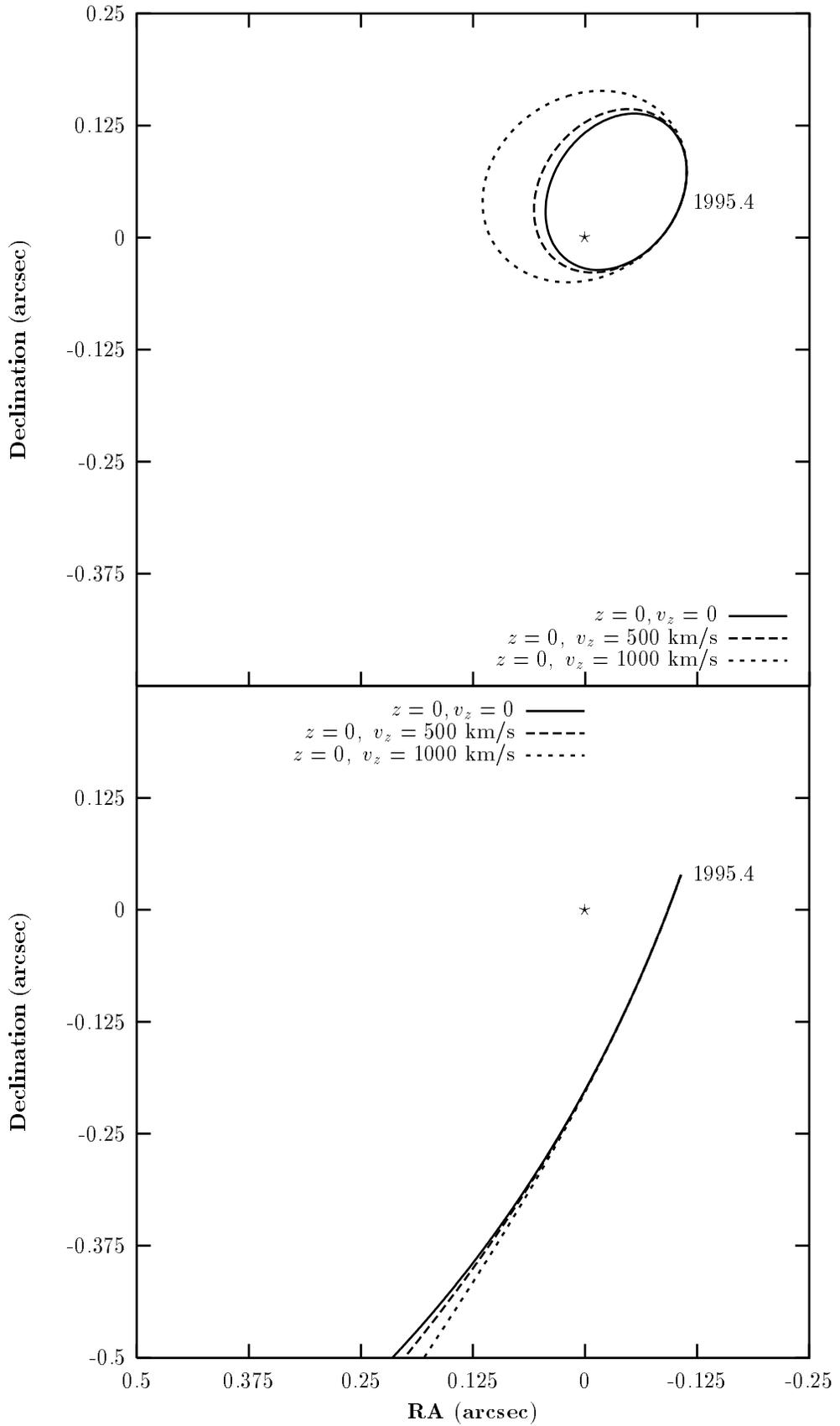
Fig.9:
Projected orbits of the star S0-1 in the case of a supermassive
BH (top panel)
and in the case of a neutrino ball (lower panel) of
$M=2.6 \times 10^{6} M_{\odot}$.
In this graph we explore how the orbits  are affected by the
uncertainty in $v_z$. The labels for the orbits are
given in the graph. Here,
 $v_x=470$ km/s, $v_y=-1330$ km/s and $z=0$.

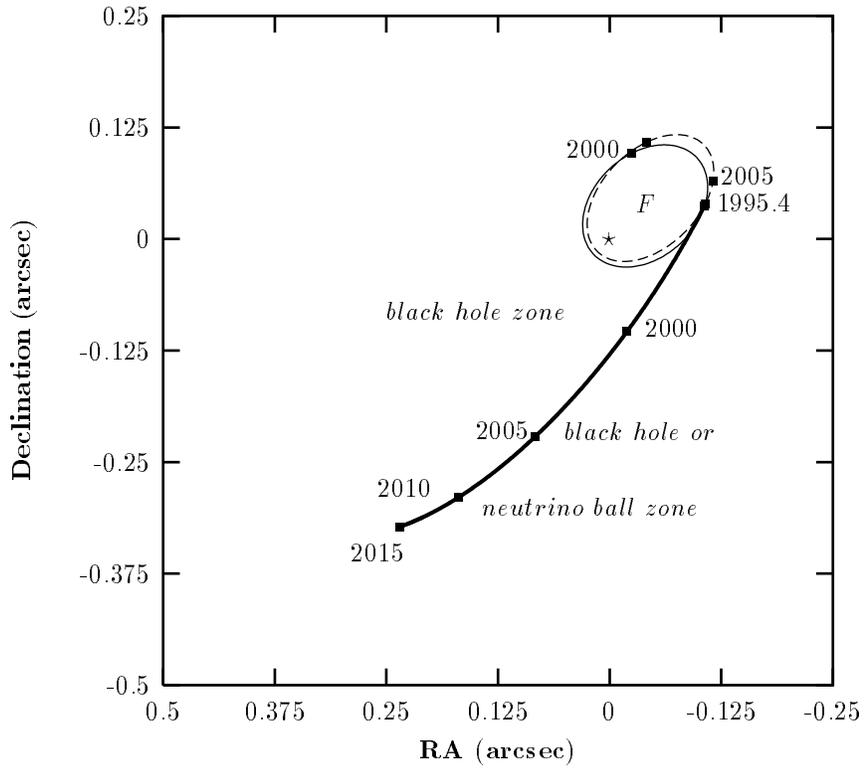
Fig.10: Prediction regions for the supermassive central object.
This graph  combines line 9 from Fig.~4 and
lines 2 and 4 from  Fig.~5.
If the star S0-1  will be found inside the ellipses (region $F$),
this will rule out  both the BH and the neutrino ball
models . If the star S0-1 will eventually
be found in the upper-left zone of the graph,
i.e. up and left 
of the thick orbit, this
will rule out the neutrino ball
interpretation  for the chosen neutrino mass.
Finally, if S0-1 will be found to the
right and below the thick line, then the supermassive
central object should be interpreted
either as a BH with large $z$ or as a neutrino ball.
\end{document}